\documentclass{article}
\usepackage[utf8]{inputenc}
\usepackage{amsmath}
\usepackage{commath}
\usepackage{graphicx}
\usepackage{float}
\usepackage{amssymb}
\begin{document}

%

% ======================================================================
% ============== User defined commands =========================================

\newcommand{\Apref}[1]{Appendix~\ref{#1}}
\newcommand{\Secref}[1]{Section~\ref{#1}}
\newcommand{\Eqref}[1]{Equation~\eqref{#1}}
\newcommand{\Figref}[1]{Figure~\ref{#1}}
\newcommand{\Tabref}[1]{Table~\ref{#1}}

\newcommand{\eVdist}{\kern-0.06em}
\newcommand{\Ev}{\text{e\eVdist V}}     % solely as unit
\newcommand{\Kev}{\text{ke\eVdist V}}
\newcommand{\Mev}{\text{Me\eVdist V}}
\newcommand{\Gev}{\text{Ge\eVdist V}}
\newcommand{\Tev}{\text{Te\eVdist V}}
\newcommand{\ev}{\:\text{e\eVdist V}}   % along with a number
\newcommand{\kev}{\:\text{ke\eVdist V}}
\newcommand{\mev}{\:\text{Me\eVdist V}}
\newcommand{\gev}{\:\text{Ge\eVdist V}}
\newcommand{\tev}{\:\text{Te\eVdist V}}
\newcommand{\s}{\:\text{s}}

\newcommand{\rep}[1]{\ensuremath\boldsymbol{#1}}
\newcommand{\crep}[1]{\ensuremath\overline{\boldsymbol{#1}}}

\newcommand{\CenterObject}[1]{\ensuremath{\vcenter{\hbox{#1}}}}
\newcommand{\CenterEps}[2][1]{\ensuremath{\vcenter{\hbox{\includegraphics[scale=#1]{#2.eps}}}}} % Input eps files - Usage: \CenterEps[ScaleFactor]{FileName}
\newcommand{\D}{\mathrm{d}}
\newcommand{\I}{\mathrm{i}}
\newcommand{\SimpleRoot}[1]{\alpha_{(#1)}}
\newcommand{\FundamentalWeight}[1]{\mu^{(#1)}}
\newcommand{\E}[1]{\ensuremath{\mathrm{E}_{#1}}} % e.g. \E{8}
\newcommand{\G}[1]{\ensuremath{\mathrm{G}_{#1}}}
\newcommand{\SO}[1]{\ensuremath{\mathrm{SO}(#1)}}
\newcommand{\SU}[1]{\ensuremath{\mathrm{SU}(#1)}}
\newcommand{\U}[1]{\ensuremath{\mathrm{U}(#1)}}
\newcommand{\Z}[1]{\ensuremath{\mathbbm{Z}_{#1}}} % z_N ->\Z{N}

\newcommand{\Y}{\ensuremath{\boldsymbol{Y}}}
\newcommand{\Yu}{\ensuremath{\boldsymbol{Y}_{\!\!u}}}
\newcommand{\Yd}{\ensuremath{\boldsymbol{Y}_{\!\!d}}}
\newcommand{\Ye}{\ensuremath{\boldsymbol{Y}_{\!\!e}}}

\newcommand{\Hu}{\ensuremath{H_u}}
\newcommand{\Hd}{\ensuremath{H_d}}
\newcommand{\qHu}{\ensuremath{q_{\Hu}}}
\newcommand{\qHd}{\ensuremath{q_{\Hd}}}
\newcommand{\singlet}{\ensuremath{N}}
\newcommand{\singletbar}{\ensuremath{\overline{N}}}
\newcommand{\dilaton}{\ensuremath{S}}
\newcommand{\tensor}{\otimes}

\newcommand{\red}[1]{\textcolor{red}{#1}}
\newcommand{\blue}[1]{\textcolor{blue}{#1}}
\newcommand{\purple}[1]{\textcolor{purple}{#1}}

\newcommand{\DHL}[0]{DHL~\cite{Dreiner:2012ae}}

\hyphenation{FCNCs}
\hyphenation{gau-gi-no}
\hyphenation{im-port-ant}
\hyphenation{coup-lings}
\hyphenation{or-bi-fold}

\def\mytitle{Revisiting Electroweak Phase Transition with Varying Yukawa Coupling Constants}

\title{\mytitle}

% ======================================================================

\begin{titlepage}

\begin{flushright}
UCI--TR--2018--10
\end{flushright}

\vspace*{1.0cm}

\begin{center}
{\large\bf\boldmath\mytitle\unboldmath}

\vspace{1cm}

\textbf{
Arianna Braconi\footnote[1]{Email: \texttt{abraconi@uci.edu}}, 
Mu--Chun Chen\footnote[2]{Email: \texttt{muchunc@uci.edu}}, 
Geoffrey Gaswint\footnote[3]{Email: \texttt{ggaswint@uci.edu}}
}
\\[5mm]
\textit{\small
{}$^a$
Department of Physics and Astronomy, University of California,\\
~~Irvine, California 92697--4575, USA
}

\end{center}

\vspace{1cm}

\begin{abstract}
We revisited the scenario of electroweak baryogenesis in the presence of large Yukawa couplings, in which it was found previously that a strongly first order electroweak phase transition can occur with the Higgs mass at its observed value of 125 GeV. Given the sensitivity of the running of the Higgs quartic coupling on the Yukawa coupling constants, we find that the addition of order one Yukawa couplings beyond the top quark drastically lowers the scale at which the Higgs potential becomes unstable. Specifically, even with only one additional order one Yukawa coupling, the scalar potential becomes unstable already at the TeV scale, assuming the Standard Model values for the Higgs sector parameters at the electroweak scale. Furthermore, by assuming the Standard Model values for the Higgs sector parameters at the TeV scale, the quartic coupling constant is driven to be larger than its Standard Model value at the electroweak scale. This in turn predicts a much lighter Higgs mass than the measured value of 125 GeV. In this scenario, the strength of the electroweak phase transition is also significantly weakened.
\end{abstract}
% mention Higgs mass lower than SM value

\end{titlepage}

\pagenumbering{arabic}

%%%%%%%%%%%%%%%%%%%%%%
%%%%%%%%%%%%%%%%%%%%%%

\section{Introduction}

The origin of the observed cosmological matter-antimatter asymmetry of the Universe remains an outstanding question in both particle physics and cosmology. Sakharov pointed out~\cite{Sakharov:1967dj} that in order for the matter-antimatter asymmetry to be generated dynamically, three conditions are required: (i) Baryon number violation, (ii) C and CP violations, and (iii) departure from thermal equilibrium. There are three possible ways to realize departure from thermal equilibrium that have been utilized in mechanisms for baryogenesis: (i) Out-of-equilibrium decay of heavy particles, (ii) electroweak phase transition, and (iii) dynamics of topological defects.

In electroweak baryogenesis, the out-of-equilibrium condition is achieved if the electroweak phase transition (EWPT) is strongly first order. This in turn requires a light Higgs mass, $m_{H} \lesssim 72$ GeV ~\cite{Grojean:2004xa} in the Standard Model (SM). Clearly this constraint is in conflict with the observed Higgs mass of $125$ GeV. In constrained Minimal Supersymmetric Standard Model (MSSM), a strongly first order EWPT requires a light Higgs which is inconsistent with observation. On the other hand, it has also been shown that a very narrow parameter space can be made consistent with the observed Higgs mass and the strong first order phase transition, by fine-tuning the supersymmetric parameters (see, e.g. ~\cite{Carena:2012np}).

Several approaches have been proposed to obtain a strongly first order EWPT while maintaining a Higgs mass consistent with  the observed value. One way is to expand the scalar sector by adding scalar singlet(s)~\cite{Espinosa:2007qk} or by adding electroweak triplet scalar fields~\cite{Garcia-Pepin:2016hvs}. It was pointed out in \cite{Berkooz:2004kx} that additional large Yukawa coupling constants in the early Universe can drive the electroweak phase transition to be strongly first order. In~\cite{Baldes:2016gaf} and ~\cite{Baldes:2016rqn} it was proposed that varying Yukawa couplings, possibly via the Froggatt-Nielsen mechanism, can also yield a strongly first order phase transition. 

In~\cite{Baldes:2016rqn} an analysis of the effective potential and phase transition with varying Yukawa couplings was carried out. The ansatz for the variation in the Yukawa couplings was taken to be 
\begin{equation}
y(\phi)=
\begin{cases}
 y_1 \left(1 - \frac{\phi}{v}\right) + y_0  &\quad  0 \leq \phi \leq v \\
 y_0  &\quad v \leq \phi \; ,
\end{cases}
\end{equation} 
with $\phi$ being the value of the Higgs field, and $y_1$ and $y_0$ are constants. The same ansatz is used in this analysis. There are three effects of the large Yukawa couplings during the electroweak phase transition: (i) In the $T=0$ one-loop corrections, the large Yukawa couplings result in the lowering of the scalar potential in the region of $0 < \bigl< \phi \bigr> < v$ thus weakening the phase transition, (ii) The finite temperature one-loop correction from the fermions adds to the potential and strengthens the phase transition, (iii) The large Yukawa coupling present in the Higgs daisy correction also strengthens the phase transition. 

Large Yukawa couplings also have a substantial effect on the running of the Higgs quartic coupling. The $\beta$ function for the quartic coupling is sensitive to changes in the Yukawa couplings, significantly so for large ones. When parameters in the Higgs sector are kept to be the Standard Model values at the weak scale, increasing the Yukawa couplings to be of order one lowers the scale at which the quartic coupling becomes negative, thus also lowering the scale at which the Higgs potential becomes unstable. If the value of the quartic coupling in the UV is kept at the Standard Model value, the renormalization group corrections driven by large Yukawa couplings  then lead to the quartic coupling value that is higher than the Standard Model value at the weak scale. Consequently, this results in a low Higgs mass that is in conflict with observation. 

The paper is organized as follows: in Sec.~\ref{sec:Veff}, we review the calculation of the effective scalar potential at finite temperature. This is then followed by Sec.~\ref{sec:stability} where we study the renormalization group evolution of the coupling constants and the stability of the scalar potential.  
In Sec.~\ref{sec:ewpt}, we present the analysis of the electroweak phase transition by including the effects of the renormalization group corrections. Sec.~\ref{sec:conclude} concludes the paper.

\section{Review of Effective Potential Calculation} 
\label{sec:Veff}

The calculation of the effective potential was developed~\cite{Coleman:1973jx,Jackiw:1974cv} using the path integral formalism. For a review, see, for example,~\cite{Quiros:1999jp}. The effective scalar potential is given by the tree level potential, plus the one-loop zero temperature correction, the one-loop finite temperature correction, and the daisy correction. 
\begin{equation}
V_{\textbf{eff}} (\phi, T) = V_{0} (\phi) + V_{1}^{T=0} (\phi) +  V_{1}^{T\neq 0}  (\phi, \, T) + V_{\text{Daisy}} (\phi, \, T)\; \; .
\end{equation}
Both the zero-temperature and finite-temperature one loop corrections receive contributions from all particles coupled to the Higgs, while the daisy correction receives contributions only from the bosons. 

%Tree Level Potential
The tree level potential, $V_{0}$, used in this calculation is given by
\begin{equation}
V_{0} = -\mu^{2}\abs{H}^2 + \lambda \abs{H}^4 \; ,
\end{equation}
with $\mu^2 > 0$ and electroweak symmetry is broken by the usual Higgs doublet 
\begin{equation}
H = (0, \frac{\phi}{\sqrt{2}}) \; . 
\end{equation}

%One Loop Zero Temperature 
The one-loop contribution at zero temperature is obtained by summing over all  1PI diagrams with one loop and zero external momenta. After renormalization using the $\overline{\text{MS}}$ scheme, the one loop zero temperature contribution is given by 
\begin{equation}
V_{1}^{T=0}(\phi) = \sum\limits_i \frac{n_i (-1)^{F_i}}{64 \pi^2} \left[m_i^4(\phi) \left(\log\left[\frac{m_i^2(\phi)}{m_i^2(v)} \right] - \frac{3}{2} \right) + 2 m_i^2(\phi) m_i^2(v)\right] \; ,
\end{equation}
where the sum is over all particles that couple to the Higgs. $F$ is 0 or 1 for bosons and fermions, respectively, and $n_i$ is the particle's multiplicity (1 for each boson, 4 for colorless Dirac fermions, and 12 if the fermion also carries color). $V_1^{T=0}(\phi)$ already includes all counter-terms necessary to maintain the tree-level values of the mass and coupling constant.

%One Loop Finite Temperature 
To calculate the finite temperature contributions, it is necessary to use the finite temperature Feynman rules and to sum over the Matsubara frequencies. The result is given by 
\begin{equation}
V_1^{T\neq 0} (\phi, T)  =  \sum\limits_i \frac{n_i (-1)^{F_i}T^4 }{2\pi^2} \, J_{\text{B/F}}\left(\frac{m(\phi)^2}{T^2} \right) \; ,
\end{equation}
where $J_{B/F}$ is the thermal bosonic/fermionic function
\begin{equation}
J_{B/F} = \int \limits_0^{\infty} dx \, x^2 \log\left[1 - (-1)^F e^{-\sqrt{x^2+m^2/T^2}}\right] \; .
\end{equation}
This integral has convenient high temperature approximations for $m^2/T^2 \ll 1$, 
\begin{align}
J_{\text{B}}\left(\frac{m^2}{T^2}\right) &= -\frac{\pi^4}{45}+\frac{\pi^2}{12}\frac{m^2}{T^2}-\frac{\pi}{6}\left(\frac{m^2}{T^2}\right)^{3/2}-\frac{1}{32}\frac{m^4}{T^4}\log\left(\frac{m^2}{a_b T^2}\right) \; ,\\
J_{\text{F}}\left(\frac{m^2}{T^2}\right) &= \frac{7\pi^4}{360}-\frac{\pi^2}{24}\frac{m^2}{T^2}-\frac{1}{32}\frac{m^4}{T^4}\log\left(\frac{m^2}{a_f T^2}\right) \; .
\end{align}
But in this analysis the thermal bosonic and fermionic functions were evaluated numerically, since near the critical temperature the condition that $m^2/T^2 \ll  1$ is violated. As observed in~\cite{Baldes:2016rqn}, if varying Yukawa couplings are assumed, the fermion contribution can produce a cubic term in $\phi$, normally only present in the bosonic contribution. This has the effect of strengthening the phase transition and can allow for a first-order  transition. 

%Daisy
The one-loop approximation is valid at temperatures below the critical temperature; however, near the critical temperature this approximation breaks down. Quadratically divergent bubbles that add a factor of $\lambda T^2 / m^2$ to the two-point functions can be safely ignored at low temperatures where $\lambda T^2/m^2 \ll 1$.  At and above the critical temperature, these diagrams (daisy, or ring diagrams) must be accounted for by resumming over an infinite number of diagrams at every order. This is equivalent to replacing the particle mass by an effective mass, $m^2 \rightarrow m^2 + \Pi(T)$, where $\Pi(T)$ is the thermal mass of the particle.
The daisy contribution is given by 
\begin{equation}
V_{\text{Daisy}}(\phi, T) = \sum\limits_i \frac{n_i T}{12 \pi} \left(m_i^3(\phi) - \left[m_i^2(\phi)+\Pi_i(T)\right]^{3/2}\right) \; .
\end{equation}
Here the $n_i$ is taken to be 1 for the Z and Higgs Bosons, and 2 for the W. In the case of the Higgs (with constant Yukawa couplings) the thermal mass calculated in~\cite{Katz:2014bha} is used
\begin{equation}
\Pi_{\phi}(T) = \left(\frac{3}{16} g_2^2 + \frac{1}{16} g_y^2 + \frac{\lambda}{2} + \frac{y_t^2}{4}\right)T^2 \; .
\end{equation}
And with varying Yukawa couplings, 
\begin{equation}
\Pi_{\phi}(\phi, T) = \left(\frac{3}{16} g_2^2 + \frac{1}{16} g_y^2 + \frac{\lambda}{2} + \frac{y_t^2}{4} +\frac{n_{*} \, y(\phi)^2}{48}\right)T^2 \; .
\end{equation}
Here $n_{*}$ is taken to be 60 to account for the additional five quarks, $g_2$ and $g_y$ are the weak and hypercharge coupling constants. 

\section{Stability of Theory with Additional Yukawa Couplings of order Unity}
\label{sec:stability}
In the Standard Model, the Higgs quartic coupling becomes negative around $10^{10}$ GeV, rendering the Higgs potential unstable.  Increasing the number of Yukawa couplings that are of order one drastically lowers the scale at which the quartic coupling becomes negative. This can be seen analytically from the 1-loop $\beta$ function (equation \ref{eq:quartic} in appendix A). If we assume only one large quark Yukawa coupling and no CKM mixing, then the leading order terms can be approximated as
\begin{equation}
\beta_{\lambda}^{(1)} \approx 24 \lambda^{2}  + 12 \lambda y^{2}  - 6 y^{4} \; . 
\end{equation}
If additional quark Yukawa couplings of order 1 are added, then the $\beta$ function becomes
\begin{equation}
\beta_{\lambda}^{(1)} \approx 24 \lambda^{2}  + 12 \lambda n y^{2}  - 6 n y^{4} \; ,
\end{equation} 
where $n$ is the number of order one Yukawa couplings. The addition of these large Yukawa couplings has the effect of driving the Higgs quartic coupling negative at a lower scale.

\begin{figure}[H]
\includegraphics[width=1.0\linewidth]{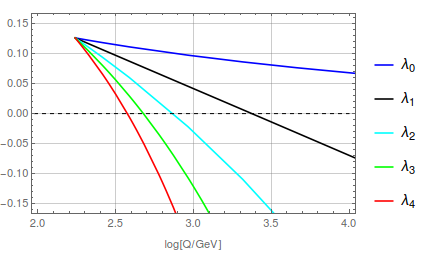}
\caption{RGE running of the Higgs quartic coupling when adding additional Yukawa couplings equal to 1. The subscript number in the legend represents how many additional Yukawa couplings were added. The subscript 0 corresponds to the case of the Standard Model.}
\label{fig:upward_instability}
\end{figure}

To evaluate the full renormalization group equations (RGEs) with large Yukawa couplings, and their effect on the running of the Higgs quartic coupling, the Mathematica package SARAH~\cite{Staub:2015kfa} was used. In the first case, Yukawa couplings equal to one were successively added to the Standard Model at the electroweak scale. The top quark Yukawa coupling and all other parameters were left as their Standard Model values. The results can be seen in Figure \ref{fig:upward_instability},  where it is clear that this change drastically lowers the scale at which the Higgs quartic coupling becomes negative, pushing it very close to the electroweak scale. 

\begin{figure}[H]
\includegraphics[width=1.0\linewidth]{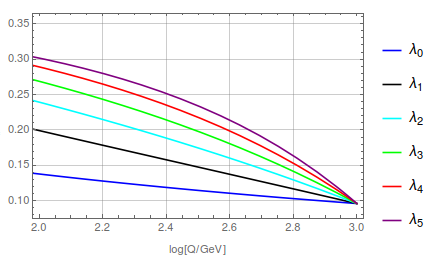}
\caption{RGE running of the Higgs quartic coupling when adding additional Yukawa couplings equal to 1. The subscript number in the legend represents how many additional Yukawa couplings were added. The subscript 0 corresponds to the case of the Standard Model.}
\label{fig:downward_instability_1}
\end{figure}

\begin{figure}[H]
\includegraphics[width=1.0\linewidth]{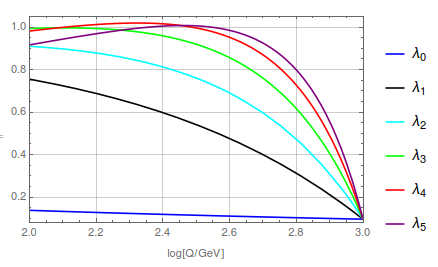}
\caption{RGE running of the Higgs quartic coupling when adding additional Yukawa couplings equal to 2. The subscript number in the legend represents how many additional Yukawa couplings were added. The subscript 0 corresponds to the case of the Standard Model.}
\label{fig:downward_instability_2}
\end{figure}

In the next case considered, the order one quark Yukawa couplings are imposed at 1 TeV. Figure \ref{fig:downward_instability_1} has the Yukawa couplings set equal to 1, while Figure \ref{fig:downward_instability_2} has the Yukawa couplings equal to 2. As shown in the figures, adding Yukawa couplings of order one or greater makes the Higgs quartic coupling larger than the observed value at the electroweak scale. This in turn has the effect of lowering the Higgs mass, as seen in the following section.

\section{Renormalization Group Improved Effective Potential and Electroweak Phase Transition}
\label{sec:ewpt}

\begin{figure}[t!]
\centering 
\includegraphics[width=0.8\linewidth]{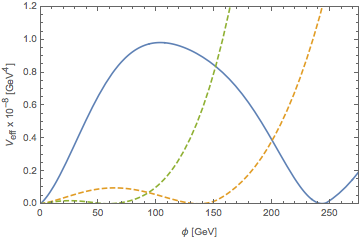}
\caption{Solid line: The effective potential with varying Yukawa couplings $y_1=2$, and Standard Model values of the coupling constants in the Higgs sector. Dashed lines: The effective potential with varying Yukawa couplings and RGE-improved values of the  Higgs sector coupling constants. The orange curve has $y_1=1$; the green curve has $y_1=2$. } 
\label{fig:all_potentials}
\end{figure}

At the electroweak scale, the coupling constants of the Higgs potential are taken as $\mu = 89$ and $\lambda = 0.13$. This allows for a first order phase transition with varying Yukawa couplings at 115 GeV, and shown by the blue solid line in Figure \ref{fig:all_potentials}.  However, this does not take into account the effects that additional large Yukawa couplings would have on the running of the RGEs and therefore on the values of the coupling constants at that scale. 

\begin{figure}[h]
\centering 
\includegraphics[width=0.8\linewidth]{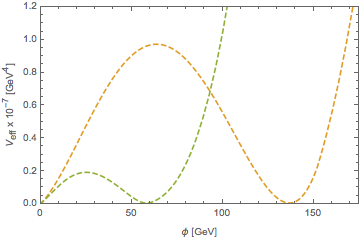}
\caption{The effective potential with RGE-improved values. The orange curve has $y_1 =1$,  $\lambda = 0.27$ and $\mu = 79$.  The green curve has $y_1= 2$,  $\lambda = 0.97$ and $\mu = 59$. } 
\label{fig:2_potentials}
\end{figure}

To account for these additional Yukawa couplings, the RGEs were run downwards from the TeV scale, with a successively increased  number of additional large Yukawa couplings. The potential of the two most extreme cases are shown, where the top Yukawa coupling retains its Standard Model value but the five additional quark Yukawa couplings are set equal to one and two, corresponding to the orange and green curves in Figures~\ref{fig:all_potentials} and~\ref{fig:2_potentials}, respectively.

The large Yukawa couplings increase the value of the quartic coupling at the electroweak scale, which in turn lowers both the vev and the temperature at which a phase transition occurs. For a transition to be considered strongly first order, it must meet the condition that $\phi_c/T_c  \gtrsim 1.3$~\cite{Quiros:1999jp}. In the case of the additional five quark Yukawa couplings equal to 1, shown by the orange curve in Figure~\ref{fig:2_potentials}, the quartic coupling increases to 0.27 which lowers the vev to 153 GeV, leading to a Higgs mass of 112 GeV.  In this scenario the critical temperature decreases to 110 GeV, and the phase transition is still strongly first order with $\phi_c/T_c = 1.39$. However, in the case of additional Yukawa couplings equal to 2, shown by the green curve in Figure~\ref{fig:2_potentials}, the quartic coupling increases to 0.97, the vev decreases to only 59 GeV, and the critical temperature is only 51 GeV. Here the phase transition is no longer strongly first order with $\phi_c/T_c = 1.16$. The predicted Higgs mass in this case is also drastically lowered, to only 54 GeV. 

\section{Conclusion}
\label{sec:conclude}

% Remove time from time varying Yukawa
% Capitalize Standard Model
% Scalar potential becomes unstable

Requiring that the electroweak phase transition in the Standard Model be strongly first order constrains the Higgs mass to be $m_{H} \lesssim 72$ GeV. With an observed Higgs mass of $125$ GeV, it is clear that a strongly first order electroweak phase transition requires physics beyond the Standard Model. A new scenario was proposed in reference \cite{Berkooz:2004kx,Baldes:2016gaf} which introduces the idea that large Yukawa couplings could cause a strongly first order electroweak phase transition. 

The Higgs quartic coupling in the Standard Model is positive up to around $10^{10}$ GeV, then the quartic coupling turns negative, making the Higgs scalar potential unstable. If additional Yukawa couplings of order 1 are present during the electroweak phase transition, then the Higgs scalar potential will become unstable at a lower scale than in the case of the Standard Model. If additional Yukawa couplings of order 1 are present at the TeV scale, then the Higgs quartic coupling is driven to be larger at the electroweak scale than its Standard Model value. This in turn predicts a lighter Higgs than the measured value. In addition to the inconsistency with the observed Higgs mass, there also exist severe constraints~\cite{Lillard:2018zts} from cosmology on the proposed scenario with varying Yukawa couplings. All together, these limitations render this simplest setup with large varying Yukawa couplings not a viable mechanism for baryogenesis. 

\section*{Acknowledgements}
We would like to thank Michael Ratz for posing the question which led to the project and for discussions. We would also like to thank Iason Baldes for helpful communications. The work of A.B. is supported, in part, by a GAANN Fellowship. The work of M.-C.C. is supported, in part, by the National Science Foundation under Grant No. PHY-1620638. M.-C.C. would like to thank the hospitality of the Aspen Center for Physics, which is supported by National Science Foundation grant PHY-1607611, where part of this work was performed.

\appendix

\section{Renormalization Group Equations}
\label{sec:RGE}

Appendix A shows the renormalization group equations (RGEs) at one loop that are used in our analysis. The RGEs for the Higgs quartic coupling, and scalar mass at one loop for the Standard Model~\cite{Arason:1991ic} are

\begin{eqnarray} 
\frac{d\lambda}{d t} & = & \frac{1}{16\pi^{2}} \beta_{\lambda}^{(1)} \; , 
\\
\frac{d\mu}{d t} & = & \frac{1}{16\pi^{2}} \beta_{\mu}^{(1)} \; ,
\end{eqnarray}
where $t = \ln E$, where E is the running scale. The beta-function coefficients are given by

%lambda -> 2*lambda retrieves sm_parameters rge reference in google drive
\begin{eqnarray}
\beta_{\lambda}^{(1)} & = & 24 \lambda^{2} - \frac{9}{5} g_{1}^{2}\lambda - 9 g_{2}^{2}\lambda  + \frac{9}{8}\biggl( \frac{3}{25} g_{1}^{4} + \frac{2}{5} g_{1}^2 g_{2}^{2} +g_{2}^{4} \biggr) \;
\label{eq:quartic}
\\ \nonumber
&& + 4 \lambda \mbox{Tr}\Big({3 Y_{u}^{\dagger} Y_{u} + 3 Y_{d}^{\dagger}Y_{d} + Y_{e}^{\dagger}Y_{e}}\Big)
\\ \nonumber
&& - 2 \mbox{Tr}\Big({3 (Y_{u}^{\dagger} Y_{u})^{2} + 3 (Y_{d}^{\dagger}Y_{d})^{2} + (Y_{e}^{\dagger}Y_{e})^{2}}\Big)
\; , \\
\beta_{\mu}^{(1)} & = & 12 \mu \lambda +
2 \mu \mbox{Tr}\Big({Y_e  Y_{e}^{\dagger}}\Big)  + 6 \mu \mbox{Tr}\Big({Y_d  Y_{d}^{\dagger}}\Big)  + 6 \mu \mbox{Tr}\Big({Y_u  Y_{u}^{\dagger}}\Big)
\\ \nonumber
&& -\frac{9}{10} g_{1}^{2} \mu  -\frac{9}{2} g_{2}^{2} \mu  \; .
\\ \nonumber
\label{eq:higgs_mass}
\end{eqnarray}
The gauge couplings for the Standard Model at the one loop level are

%Gauge Couplings RGEs
\begin{eqnarray}
\frac{dg_{i}}{dt} & = & \frac{1}{16\pi^{2}}\beta_{g_{i}}^{(1)} \; ,
\end{eqnarray}
where the three $\beta_{g_{i}}$ functions are expressed as
\begin{eqnarray}
\beta_{g_1}^{(1)} & = & 
\frac{41}{10} g_{1}^{3} \; , \\ 
\beta_{g_2}^{(1)} & = & 
-\frac{19}{6} g_{2}^{3} \; , \\ 
\beta_{g_3}^{(1)} & = & 
-7 g_{3}^{3} \; .
\end{eqnarray}
The RGE's for the Yukawa couplings in the Standard Model at one loop are summarized as

%Up family RGE
\begin{eqnarray}
\beta_{Y_u}^{(1)} & = &  
-\frac{3}{2} \Big(- {Y_u  Y_{u}^{\dagger}  Y_u}  + {Y_u  Y_{d}^{\dagger}  Y_d}\Big) 
+Y_u \Big( 3 \mbox{Tr}\Big({Y_d  Y_{d}^{\dagger}}\Big)  + 3 \mbox{Tr}\Big({Y_u  Y_{u}^{\dagger}}\Big)
\\
&&
-8 g_{3}^{2}  -\frac{17}{20} g_{1}^{2}  -\frac{9}{4} g_{2}^{2}  + \mbox{Tr}\Big({Y_e  Y_{e}^{\dagger}}\Big)\Big) \; ,
\nonumber\\
\beta_{Y_d}^{(1)} & = & 
\frac{1}{4} \Big(6 \Big(- {Y_d  Y_{u}^{\dagger}  Y_u}  + {Y_d  Y_{d}^{\dagger}  Y_d}\Big) - Y_d \Big(-12 \mbox{Tr}\Big({Y_d  Y_{d}^{\dagger}}\Big) 
\\
&& -12 \mbox{Tr}\Big({Y_u  Y_{u}^{\dagger}}\Big) + 32 g_{3}^{2}  -4 \mbox{Tr}\Big({Y_e  Y_{e}^{\dagger}}\Big)  + 9 g_{2}^{2}  + g_{1}^{2}\Big)\Big) \nonumber \; ,
\\
\beta_{Y_e}^{(1)} & = &
\frac{3}{2} {Y_e  Y_{e}^{\dagger}  Y_e}  + Y_e \Big(3 \mbox{Tr}\Big({Y_d  Y_{d}^{\dagger}}\Big)  + 3 \mbox{Tr}\Big({Y_u  Y_{u}^{\dagger}}\Big) -\frac{9}{4} g_{1}^{2}  -\frac{9}{4} g_{2}^{2}
\\
&& + \mbox{Tr}\Big({Y_e  Y_{e}^{\dagger}}\Big)\Big) \nonumber \; .
\end{eqnarray}

\bibliography{PT}
\addcontentsline{toc}{section}{Bibliography}
\bibliographystyle{NewArXiv} 
\end{document}